\documentclass[preprint,amsmath,amssymb]{revtex4}
\usepackage{graphicx}
\usepackage{dcolumn}
\usepackage{bm}

\begin{document}
\title{Diffusion coefficient and DC conductivity of anisotropic static black hole   }
\author{  Z.~Amoozad $^{a}$ \footnote{Electronic
address:~z.amoozad@stu.umz.ac.ir }
and  J. ~Sadeghi $^a$ \footnote{Electronic address: ~pouriya@ipm.ir , corresponding author } }
\address{$^{a}$ Sciences Faculty, Department of Physics, University of Mazandaran, Babolsar, Iran \\
P.O.Box 47416-95447.}

\begin{abstract}
In this study we apply two different methods in the context of $AdS/CFT$ correspondence and calculate the diffusion coefficient and $DC$ conductivity of a four-dimensional spatially anisotropic static black hole. First, the \emph{modified} transport coefficients is obtained by stretched horizon method and Fick's law in the context of the membrane paradigm. In order to do such calculation, we use the Maxwell equations with electromagnetic gauge field propagating in two dimensions. Two dimensional propagating gauge field leads to the complex transport coefficients which is proved by present paper.
In second step, we explain electro-thermal method and employ an effective vector field and extract retarded Green's function on the classical boundary. Then, $DC$ conductivity and diffusion coefficient are obtained by using Kubo formula. Our calculation can be applied on two well-known examples of anisotropic black holes as the Einstein-Maxwell-dilaton-axion model and AdS-Einstein-Maxwell-dilaton-axion in massive gravity.           \\\\
{\bf Keywords:} Anisotropic black hole; Diffusion coefficient; $DC$ conductivity.
\end{abstract}
\maketitle

\section{Introduction}
\label{intro}
As we know for the strongly interacting systems such as condensed matter systems we can not use traditional perturbation method. A new technique which is useful and solve the problem of strongly interacting systems is $AdS/CFT$ correspondence~\cite{a,b,c,d}.  The nonperturbativ properties of strongly interacting systems such as scaling behavior of transport coefficients is resolved by holography~\cite{e,f,g,h}. But for the charged black holes, because of the existence of a traditional symmetry, working on transport coefficient with this new method 
leads to the unexpected behavior of electric conductivity, namely, it exhibits a delta function behavior in the zero frequency limit.

There are different methods which break translational symmetry and achieve a finite conductivity .  such as lattice structure of AdS geometry~\cite{i,j,k}, lattice structure which imposes periodic boundary condition on the chemical potential~\cite{l,m,n,o}, massive gravity theory~\cite{p,q,r,s} and introducing additional fields depending on spatial coordinates~\cite{t,u}. In these methods, spatial anisotropy can be extracted when the rotational symmetry of the system is breaking. For the last technique, the gravity has local gauge field, dilaton and axion fields. In the $AdS/CFT$ context, the local bulk gauge field will be identified by the matter with corresponding global symmetry while dilaton field is mapped into the coupling constant of the dual theory.

Different calculations have been done on anisotropic black branes. The anisotropic plasma, its thermodynamics and instabilities have been investigated in~\cite{A1,A2,A3,A4}. Anisotropic black branes in higher curvature gravity and holographic renormalization have been studied in~\cite{A5}. Some authors have analysed heavy ions collisions~\cite{A6}, hyperscaling violation~\cite{A7} and anisotropic black branes with Lifshitz scaling~\cite{A8}. The thermal diffusivity and butterfly velocity in anisotropic Q-lattice model have also been explained in~\cite{A9}.

Many works on calculating transport coefficients of black holes have been done in the literature. Some of that, lay in the membrane paradigm which the diffusion coefficient is obtained by spacetime properties for static and rotating black holes~\cite{v,w,cc}. Considering fluctuations around static black brane solutions, the diffusion relations and shear flow can be obtained in the holographically dual theory. This method is one of the examples which describes $AdS/CFT$ duality in long wavelength regime which is known as fluid/gravity duality~\cite{x,y,z,aa,bb}. Another method is the electro-thermal, in which the diffusion constant can be extracted when the electrical conductivity and susceptibility of the system are obtained~\cite{dd,ee,ff}.
Then, we introduce two known examples for the spatially anisotropic black holes which are Einstein-Maxwell-dilaton-axion model and AdS-Einstein-Maxwell-dilaton-axion in massive gravity which have been introduced in~\cite{hh,ii}. All discussed above gives us motivation to extract diffusion constant and DC conductivity of the four-dimensional spatially anisotropic static black holes.

 Here, the anisotropy will be on the spatial coordinates and corresponding parameters are only $r$-dependance. By using external vector gauge field which propagates in two dimensions $x,y$ we investigate  fluctuations of bulk metric. Choosing two dimensional fluctuation causes some difficulty on the intermediate steps, so some constraints will be added to set the stage and simplify continuation. In fact, the results are modified because they are complex while in the case where the gauge field propagates in one dimension, it yields real diffusion coefficient. Since the black hole is anisotropic, two nonzero components of diffusivity and conductivity matrix have been calculated. After that, by taking appropriate perturbation we explain electro-thermal method and obtain retarded Green's function on the classical boundary. Then, using Kubo formula, the $DC$ conductivity of system will be obtained. At final step the susceptibility of the corresponding system led us to investigate the diffusion coefficient.

We organise the paper as follows. In section II we explain general properties of static anisotropic four-dimensional black hole and present appropriate Maxwell equations and Bianchi identities. In section III, after introducing a vector gauge field and some constraints we use the membrane paradigm to extract Fick's first law. It gives a complex diffusion coefficient and DC conductivity. In section IV, the $DC$ conductivity is obtained by extracting the corresponding retarded Green's function. In that case, the susceptibility gives the diffusion constant. Then, we have devoted two examples concerning the anisotropic black holes and explain some of their properties. In the final section, some concluding remarks and notes are mentioned.

\section{ The general properties of static anisotropic four-dimensional black hole }
\label{sec:1}
The most applicable solutions in the context of the $AdS/CFT$ are black holes which are static and asymptotically $AdS$. These black holes incorporate in the $AdS/CFT$ duality to introduce some solutions to the strongly coupled field theory and condensed matter systems. Obtaining finite DC conductivity in the strongly correlated systems requires breaking translational symmetry, it leads to the anisotropic spacetimes. The importance of these spacetimes give us motivation to write general form of anisotropic $AdS$ spacetime as bellow,
\begin{equation}
\label{eq:1}
ds^2=g_{tt}(r) dt^2 + g_{rr}(r) dr^2+  g_{xx}(r) dx^2+  g_{yy}(r) dy^2.
\end{equation}
The corresponding event horizon placed at $r=r_0$ where $g_{rr}^{-1}(r_{0})=0$ and the Hawking temperature of this space time is $T_{H}= \frac{1}{4\pi}\frac{1}{\sqrt{g_{rr}g_{tt}}}\frac{d}{dr}g_{tt}|_{r=r_{0}}$.

By using membrane paradigm, we obtain proper region which is suitable to calculate diffusion coefficient and $DC$ conductivity. In this case, the stretched horizon is a flat space-like surface located in $r=r_{h}$, such that one can write following expression,
\begin{equation}
\label{eq:2}
r_{h}>r_{0},\;\;\;\;\;\;\;\;\;\ r_{h}-r_{0}\ll r_{0}.
\end{equation}
The unit normal vector on the corresponding hypersurface is a space-like vector. By using the constraint $\Phi=r=const,$ we have $n_{r}=\sqrt{g_{rr}}$.

The hydrodynamic theory in the dual field theory can be produced by currents and tensors, so in that case it is important to investigate the vector and tensor perturbations.
The dynamics of vector perturbations can be investigated by following Maxwell action,
\begin{equation}
\label{eq:3}
S_{gauge}\sim\int dx^{p+2}\sqrt{-g}(\frac{1}{g_{eff}^{2}}F^{\mu\nu}F_{\mu\nu}).
\end{equation}
Now by existence of an external gauge field $A_{\mu}$ as a perturbation, one can find currents on the stretched horizon $r=r_{h}$ as,
\begin{equation}
\label{eq:4}
J^{\mu}=n_{\nu}F^{\mu\nu}\mid_{r_{h}}.
\end{equation}
After rearrangement of the components of the current we have,
\begin{eqnarray}
\label{eq:5}
J^{t}&=&\frac{1}{g_{tt}\sqrt{g_{rr}}}F_{tr},;\;\;\;\;\;\;\;\;\;\;\;\;\;\;\;\;J^{x_{i}}=\frac{1}{g_{x_{i}x_{i}}\sqrt{g_{rr}}}F_{x_{i}r},
\end{eqnarray}
where the conservation equation of current is $\partial_{\mu}J^{\mu}=0$. Because of the antisymmetry properties of $F^{\mu\nu}$, we conclude $n_{\mu} J^{\mu}=0$, so it proves that the corresponding currents are parallel to the stretched horizon.

Now we search for other relations which come from Maxwell equations and Bianchi identities. First we extract equations from,
\begin{equation}
\label{eq:6}
\partial_{\mu}(\frac{1}{g_{eff}^{2}}\sqrt{-g}F^{\mu\nu})=0,
\end{equation}
For simplicity, the effective coupling  $g_{eff}$ has been taken as a constant. Thus, the components of the Maxwell equations  in the four dimensional space time $t,r,x,y$ will be,

\begin{equation}
\label{eq:7}
\partial_{r}(\sqrt{-g}g^{rr}g^{tt}F_{rt})+\partial_{x}(\sqrt{-g}g^{xx}g^{tt}F_{xt})+\partial_{y}(\sqrt{-g}g^{yy}g^{tt}F_{yt})=0,
\end{equation}
\begin{equation}
\label{eq:8}
\partial_{t}(\sqrt{-g}g^{tt}g^{rr}F_{tr})+\partial_{x}(\sqrt{-g}g^{rr}g^{xx}F_{xr})+\partial_{y}(\sqrt{-g}g^{rr}g^{yy}F_{yr})=0
\end{equation}
\begin{equation}
\label{eq:9}
\partial_{t}(\sqrt{-g}g^{tt}g^{xx}F_{tx})+\partial_{r}(\sqrt{-g}g^{rr}g^{xx}F_{rx})+
\partial_{y}(\sqrt{-g}g^{xx}g^{yy}F_{yx})=0
\end{equation}
\begin{equation}
\label{eq:10}
\partial_{t}(\sqrt{-g}g^{tt}g^{yy}F_{ty})+\partial_{r}(\sqrt{-g}g^{rr}g^{yy}F_{ry})+
\partial_{x}(\sqrt{-g}g^{yy}g^{xx}F_{xy})=0
\end{equation}

In second step we use Bianchi identities to extract other relations between components of field strength,
\begin{equation}
\label{eq:11}
F_{[\mu\nu,\lambda]}=0.
\end{equation}
It leads to the following equations in the four dimensional system $(t,r,x,y)$,
\begin{equation}
\label{eq:12}
\partial_{r}F_{xt}+\partial_{x}F_{tr}+\partial_{t}F_{rx}=0,
\end{equation}
\begin{equation}
\label{eq:13}
\partial_{r}F_{yt}+\partial_{y}F_{tr}+\partial_{t}F_{ry}=0,
\end{equation}
\begin{equation}
\label{eq:14}
\partial_{r}F_{xy}+\partial_{x}F_{yr}+\partial_{y}F_{rx}=0,
\end{equation}
\begin{equation}
\label{eq:15}
\partial_{x}F_{yt}+\partial_{y}F_{tx}+\partial_{t}F_{xy}=0,
\end{equation}

Reviewing properties of any two dimensional surface~\cite{gg} shows that every two dimensional surface can be represented by a conformally flat surface. So the appropriate field which propagates in such surface has plane wave representation. We emphasis that the field strength depends only on the conformal parameter $r$, therefore we use the following ansatz for the vector field propagating on the horizon,
\begin{equation}
\label{eq:16}
A_{\mu}=A_{\mu}(r)e^{-i\omega t+i\overrightarrow{q}.\overrightarrow{x}},
\end{equation}
where $\mu$ runs over $t,r,x,y$. We will use (16) to find transport coefficients in the next sections.

\section{The investigation  of Fick's first law, diffusion coefficient and $DC$ conductivity  }
As we know the calculation of  diffusion coefficient in the membrane paradigm is connected to the Fick's first law. To achieve that, we must relate different components of the Maxwell equations in such away to have $J^{x}=-D\partial_{x}J^{0}$. However, it requires some calculations and assumptions which help us to arrive a suitable result.

 First of all we suppose $g_{tt}g_{rr}=-1$ the assumption of spherically symmetric spacetime. Second assumption introduced in (16), it means that  the vector gauge field propagates in two $x,y$ directions,
 \begin{equation}
\label{eq:17}
A_{\mu}=A_{\mu}(r)e^{-i\omega t+iq_{x}x+iq_{y}y}.
\end{equation}
Thus the Maxwell equations will be,
\begin{equation}
\label{eq:18}
\partial_{r}(\sqrt{g_{xx}g_{yy}}F_{rt})+g^{tt}\partial_{x}(\sqrt{\frac{g_{yy}}{g_{xx}}}F_{xt})+
g^{tt}\partial_{y}(\sqrt{\frac{g_{xx}}{g_{yy}}}F_{yt})=0,
\end{equation}
\begin{equation}
\label{eq:19}
\sqrt{g_{xx}g_{yy}}\partial_{t}F_{tr}+g^{rr}\partial_{x}(\sqrt{\frac{g_{yy}}{g_{xx}}}F_{xr})+
g^{rr}\partial_{y}(\sqrt{\frac{g_{xx}}{g_{yy}}}F_{yr})=0,
\end{equation}
\begin{equation}
\label{eq:20}
g^{tt}\sqrt{\frac{g_{yy}}{g_{xx}}}\partial_{t}F_{tx}+\partial_{r}(\sqrt{\frac{g_{yy}}{g_{xx}}}g^{rr}F_{rx})+
\partial_{y}(\frac{1}{\sqrt{g_{xx}g_{yy}}}F_{yx})=0,
\end{equation}
\begin{equation}
\label{eq:21}
g^{tt}\sqrt{\frac{g_{xx}}{g_{yy}}}\partial_{t}F_{ty}+\partial_{r}(\sqrt{\frac{g_{xx}}{g_{yy}}}g^{rr}F_{ry})+
\partial_{x}(\frac{1}{\sqrt{g_{xx}g_{yy}}}F_{xy})=0.
\end{equation}
For simplicity we choose $A_{r}=0$, so one can rearrange the  Maxwell equations as,
\begin{equation}
\label{eq:22}
\partial_{r}(\sqrt{g_{xx}g_{yy}}F_{rt})+g^{tt}\sqrt{\frac{g_{yy}}{g_{xx}}}\partial_{x}F_{xt}+
g^{tt}\sqrt{\frac{g_{xx}}{g_{yy}}}\partial_{y}F_{yt}=0,
\end{equation}
\begin{equation}
\label{eq:23}
g^{tt}\partial_{t}F_{tr}+g^{xx}\partial_{x}F_{xr}+g^{yy}\partial_{y}F_{yr}=0,
\end{equation}
\begin{equation}
\label{eq:24}
g^{tt}\sqrt{\frac{g_{yy}}{g_{xx}}}\partial_{t}F_{tx}+\partial_{r}(\sqrt{\frac{g_{yy}}{g_{xx}}}g^{rr}F_{rx})+
\frac{1}{\sqrt{g_{xx}g_{yy}}}\partial_{y}F_{yx}=0,
\end{equation}
\begin{equation}
\label{eq:25}
g^{tt}\sqrt{\frac{g_{xx}}{g_{yy}}}\partial_{t}F_{ty}+\partial_{r}(\sqrt{\frac{g_{xx}}{g_{yy}}}g^{rr}F_{ry})+\frac{1}{\sqrt{g_{xx}g_{yy}}}\partial_{x}F_{xy}=0.
\end{equation}
These equations with Bianchi identities help us finding diffusion constant.

We follow  Ref. ~\cite{v} and  mix $t$-derivative of Eq. (23) with Bianchi identities (12) and (13) which takes,
\begin{equation}
\label{eq:26}
g^{tt}{\partial^2}_{t}F_{tr}+i\{g^{xx}q_{x}\partial_{r}F_{xt}+g^{yy}q_{y}\partial_{r}F_{yt}\}-\{g^{xx}{q_{x}}^{2}+g^{yy}{q_{y}}^{2}\}F_{tr}=0.
\end{equation}
By applying gauge field Eq. (17) we can rewrite above equation as,
\begin{equation}
\label{eq:27}
\{1+\frac{1}{\omega^{2}g^{tt}}(g^{xx}{q_{x}}^{2}+g^{yy}{q_{y}}^{2})\}F_{tr}-\frac{i}{\omega^{2}g^{tt}}\left\{g^{xx}q_{x}\partial_{r}F_{xt}+g^{yy}q_{y}\partial_{r}F_{yt}\right\}=0.
\end{equation}
In near horizon regime, we take $\mid\frac{1}{\omega^{2}g^{tt}}(g^{xx}{q_{x}}^{2}+g^{yy}{q_{y}}^{2})\mid\ll 1$, so  one can write,
\begin{equation}
\label{eq:28}
F_{tr}\sim\frac{1}{\omega^{2}g^{tt}}\left\{g^{xx}q_{x}\partial_{r}F_{xt}+g^{yy}q_{y}\partial_{r}F_{yt}\right\}.
\end{equation}
On the other hand, by taking $t$-derivative of Eq.(24),  applying Bianch identities (12) and (15), we arrive at,
\begin{equation}
\label{eq:29}
g^{tt}\partial_{t}^{2}F_{tx}
-\sqrt{\frac{g_{xx}}{g_{yy}}}\partial_{r}(\sqrt{\frac{g_{yy}}{g_{xx}}}g^{rr}\partial_{r}F_{xt})
-\sqrt{\frac{g_{xx}}{g_{yy}}}\partial_{r}(\sqrt{\frac{g_{yy}}{g_{xx}}}g^{rr}\partial_{x}F_{tr})
-\frac{q_{y}}{g_{yy}}(q_{x}F_{yt}+q_{y}F_{tx})=0.
\end{equation}
Also we take $t$-derivative of Eq.(25),  applying Bianchi identities (13) and (15), another useful equation will be obtain by following expression,
\begin{equation}
\label{eq:30}
g^{tt}\partial_{t}^{2}F_{ty}
-\sqrt{\frac{g_{yy}}{g_{xx}}}\partial_{r}(\sqrt{\frac{g_{xx}}{g_{yy}}}g^{rr}\partial_{r}F_{yt})
-\sqrt{\frac{g_{yy}}{g_{xx}}}\partial_{r}(\sqrt{\frac{g_{xx}}{g_{yy}}}g^{rr}\partial_{y}F_{tr})
+\frac{q_{x}}{g_{xx}}(q_{x}F_{yt}+q_{y}F_{tx})=0.
\end{equation}
Now we try to extract independent differential equations for the $F_{tx}$ and $F_{ty}$.  In order to  achieve  the corresponding result, we apply condition (28) in  Eq.(29) and then we have,
\begin{equation}
\label{eq:31}
F_{tx}-\frac{1}{\omega^{2}g^{tt}}\sqrt{\frac{g_{xx}}{g_{yy}}}\partial_{r}(\sqrt{\frac{g_{yy}}{g_{xx}}}g^{rr}\partial_{r}F_{tx})
-i\frac{q_{x}q_{y}}{\omega^{4}g^{tt}}\sqrt{\frac{g_{xx}}{g_{yy}}}\partial_{r}(\sqrt{\frac{g_{yy}}{g_{xx}}}g^{rr}\frac{g^{yy}}{g^{tt}}\partial_{r}F_{ty})
-\frac{q_{x}q_{y}}{\omega^{2}g^{tt}g_{yy}}F_{ty}=0.
\end{equation}
Now we are going to apply condition (28) in  Eq.(30), also we have,
\begin{equation}
\label{eq:32}
F_{ty}-\frac{1}{\omega^{2}g^{tt}}\sqrt{\frac{g_{yy}}{g_{xx}}}\partial_{r}(\sqrt{\frac{g_{xx}}{g_{yy}}}g^{rr}\partial_{r}F_{ty})
-i\frac{q_{x}q_{y}}{\omega^{4}g^{tt}}\sqrt{\frac{g_{yy}}{g_{xx}}}\partial_{r}(\sqrt{\frac{g_{xx}}{g_{yy}}}g^{rr}\frac{g^{xx}}{g^{tt}}\partial_{r}F_{tx})
-\frac{q_{x}q_{y}}{\omega^{2}g^{tt}g_{xx}}F_{tx}=0.
\end{equation}
Considering the assumption of spherically symmetric spacetime, i.e. $g_{tt}g_{rr}=-1$ and some manipulations one can write  two above equations as,
\begin{equation}
\label{eq:33}
\begin{split}
&F_{tx}-\frac{{g^{rr}}^{2}}{\omega^{2}}\left\{\partial_{r}\ln (\sqrt{\frac{g_{yy}}{g_{xx}}} g^{rr})\right\}\partial_{r}F_{tx}
-\frac{{g^{rr}}^{2}}{\omega^{2}}\partial_{r}^{2}F_{tx} \\
&-i\frac{q_{x}q_{y}{g^{rr}}^{3}}{\omega^{4}g_{yy}}\left\{\partial_{r}\ln (\frac{{g^{rr}}^{2}}{\sqrt{g_{yy}g_{xx}}})\right\}\partial_{r}F_{ty}
-i\frac{q_{x}q_{y}{g^{rr}}^{3}}{\omega^{4}g_{yy}}\partial_{r}^{2}F_{ty}
-\frac{q_{x}q_{y}g^{rr}}{\omega^{2}g_{yy}}F_{ty}=0,
\end{split}
\end{equation}

\begin{equation}
\label{eq:34}
\begin{split}
&F_{ty}-\frac{{g^{rr}}^{2}}{\omega^{2}}\left\{\partial_{r}\ln (\sqrt{\frac{g_{xx}}{g_{yy}}} g^{rr})\right\}\partial_{r}F_{ty}
-\frac{{g^{rr}}^{2}}{\omega^{2}}\partial_{r}^{2}F_{ty} \\
&-i\frac{q_{x}q_{y}{g^{rr}}^{3}}{\omega^{4}g_{xx}}\left\{\partial_{r}\ln (\frac{{g^{rr}}^{2}}{\sqrt{g_{yy}g_{xx}}})\right\}\partial_{r}F_{tx}
-i\frac{q_{x}q_{y}{g^{rr}}^{3}}{\omega^{4}g_{xx}}\partial_{r}^{2}F_{tx}
-\frac{q_{x}q_{y}g^{rr}}{\omega^{2}g_{xx}}F_{tx}=0.
\end{split}
\end{equation}

After some calculations the general independent differential equations for the $F_{tx}$ and $F_{ty}$ will be,
\begin{equation}
\label{eq:35}
\begin{split}
F_{tx}^{\prime\prime}
+(1-i\frac{q_{x}q_{y}g^{rr}}{\omega^{2}g_{xx}})
&\left\{\partial_{r}\ln (\sqrt{\frac{g_{yy}}{g_{xx}}} g^{rr})+i\frac{q_{x}q_{y}g^{rr}}{\omega^{2}g_{xx}}\partial_{r}\ln( \frac{{g^{rr}}^{2}}{\sqrt{g_{yy}g_{xx}}})\right\}F_{tx}^\prime \\
&-\frac{\omega^{2}}{{g^{rr}}^{2}}\left(1-(1+i)\frac{q_{x}q_{y}g^{rr}}{\omega^{2}g_{xx}}\right)F_{tx}=0,
\end{split}
\end{equation}

\begin{equation}
\label{eq:36}
\begin{split}
F_{ty}^{\prime\prime}
+(1-i\frac{q_{x}q_{y}g^{rr}}{\omega^{2}g_{yy}})
&\left\{\partial_{r}\ln (\sqrt{\frac{g_{xx}}{g_{yy}}} g^{rr})+i\frac{q_{x}q_{y}g^{rr}}{\omega^{2}g_{yy}}\partial_{r}\ln( \frac{{g^{rr}}^{2}}{\sqrt{g_{yy}g_{xx}}})\right\}F_{ty}^\prime \\
&-\frac{\omega^{2}}{{g^{rr}}^{2}}\left(1-(1+i)\frac{q_{x}q_{y}g^{rr}}{\omega^{2}g_{yy}}\right)F_{ty}=0.
\end{split}
\end{equation}

These differential equations are  exactly radial parts of $F_{tx}$ and $F_{ty}$  and they have complex solutions. The complexity comes from taking the gauge field which propagates in two directions $x,y$. If we take $q_y=0$, the solution of equations will be real which consistent with Ref.~\cite{v}. So for each anisotropic static spherically symmetric black hole with constraint (28) the perturbed component of vector gauge fields $F_{tx}(r)$ and $F_{ty}(r)$ will be extracted as (35) and (36).  We note that  the explicit solution to the above differential equation will be found when the metric components be identified. Here, as the perturbed gauge field $A_{\mu}=A_{\mu}(r)e^{-i\omega t+iq_{x}x+iq_{y}y}$, one can obtain,

\begin{equation}
\label{eq:37}
\partial_{r}F_{tx}(r,t,x,y)\propto (\textit{A+iB}) \partial_{t}F_{tx}(r,t,x,y),
\end{equation}

\begin{equation}
\label{eq:38}
\partial_{r}F_{ty}(r,t,x,y)\propto (\textit{M+iN}) \partial_{t}F_{ty}(r,t,x,y).
\end{equation}
We emphasis on the complexity of the coefficients,  because the solutions of Eqs. (35) and (36) are complex.

At the other hand,  we can take the Bianchi identity (12) and condition (28), so we can write,
\begin{equation}
\label{eq:39}
\partial_{r}F_{xt}+\partial_{t}F_{rx}
+\partial_{x}\left[\frac{1}{\omega^{2}g^{tt}}(g^{xx}q_{x}\partial_{r}F_{xt}+g^{yy}q_{y}\partial_{r}F_{yt})\right]=0,
\end{equation}
it is obvious that the first term in the bracket can be neglected in our regime. Again,  the Eqs (13) and (28) lead us to have following equation,
\begin{equation}
\label{eq:40}
\partial_{r}F_{yt}+\partial_{t}F_{ry}
+\partial_{y}\left[\frac{1}{\omega^{2}g^{tt}}(g^{xx}q_{x}\partial_{r}F_{xt}+g^{yy}q_{y}\partial_{r}F_{yt})\right]=0,
\end{equation}
our regime imposes that the second term on the bracket will be neglected with compere  to the first term of equation. At the other hand by applying relation (37) in the (39), we will arrive at,
\begin{equation}
\label{eq:41}
\partial_{t}\left[(\textit{A+iB})F_{tx}+F_{rx}\right]+\frac{q_{y}g^{yy}}{\omega^2 g^{tt}}\partial_{x}\partial_{r}F_{yt}=0.
\end{equation}
 One of the solutions of the above differential equations will be obtained when the first and second terms of equation be zero. Therefore, in order to have finite solution  at $t\rightarrow\infty$ it must be zero. At the other hand considering relation (38) and (40) yield,
\begin{equation}
\label{eq:42}
\partial_{t}\left[(\textit{M+iN})F_{ty}+F_{ry}\right]+\frac{q_{x}g^{xx}}{\omega^2 g^{tt}}\partial_{y}\partial_{r}F_{xt}=0,
\end{equation}
by the same reason we can take the expression in the bracket to be zero. This result shows the connection between electric and magnetic field on the horizon.

Now we try to find $A_{t}$ which will be required later. By using method of Ref.~\cite{v} in the limit $\frac{q_{x}^2}{T^{2}}\ll1$ and $\frac{q_{y}^2}{T^{2}}\ll1$ the first term in the expansion of $A_{t}$ is,
\begin{equation}
\label{eq:43}
A_{t}^{(0)}(t,x,r)=C_{0}(t)e^{iqx}\int_{r}^{\infty} \frac{dr'}{\sqrt{g_{xx}(r')g_{yy}(r')}}.
\end{equation}
In the near horizon limit the ratio of $A_{t}$ and $F_{tr}$ is a constant which in terms of metric component is,
\begin{equation}
\label{eq:44}
\frac{A_{t}}{F_{tr}}\mid_{r\approx r_{0}}=\sqrt{g_{xx}(r_{0})g_{yy}(r_{0})}\int_{r_{0}}^{\infty} \frac{dr}{\sqrt{g_{xx}(r)g_{yy}(r)}}.
\end{equation}
In addition to the above condition,  we suppose $\mid\partial_{t}A_{x}\mid\ll\mid\partial_{x}A_{t}\mid$.

Now, all requirements for derivation of Fick's first law are exist,  so we find the current as,
\begin{equation}
\label{eq:45}
\begin{split}
J^{x}&=\frac{1}{g_{xx}\sqrt{g_{rr}}}F_{xr}=-(\textit{A+iB}){g_{xx}\sqrt{g_{rr}}} F_{tx},\\
     &=(\textit{A+iB}){g_{xx}\sqrt{g_{rr}}} \partial_{x}A_{t}, \\
     &=(\textit{A+iB}){g_{xx}\sqrt{g_{rr}}} \left(\frac{A_{t}}{F_{tr}}\right)\partial_{x}F_{tr}, \\
     &=(\textit{A+iB}){g_{xx}g_{rr}} \left(\frac{A_{t}}{F_{tr}}\right)\partial_{x}J^{t}, \\
     &=-\left[-(\textit{A+iB}){g_{xx}g_{rr}} \sqrt{g_{xx}(r_{0})g_{yy}(r_{0})}\int_{r_{0}}^{\infty} \frac{dr}{\sqrt{g_{xx}(r)g_{yy}(r)}}\right]\partial_{x}J^{t}.
\end{split}
\end{equation}
and
\begin{equation}
\label{eq:46}
\begin{split}
J^{y}&=\frac{1}{g_{yy}\sqrt{g_{rr}}}F_{yr}=-(\textit{M+iN}){g_{yy}\sqrt{g_{rr}}} F_{ty},\\
     &=(\textit{M+iN}){g_{yy}\sqrt{g_{rr}}} \partial_{y}A_{t}, \\
     &=(\textit{M+iN}){g_{yy}\sqrt{g_{rr}}} \left(\frac{A_{t}}{F_{tr}}\right)\partial_{y}F_{tr}, \\
     &=(\textit{M+iN}){g_{yy}g_{rr}} \left(\frac{A_{t}}{F_{tr}}\right)\partial_{y}J^{t}, \\
     &=-\left[-(\textit{M+iN}){g_{yy}g_{rr}} \sqrt{g_{yy}(r_{0})g_{yy}(r_{0})}\int_{r_{0}}^{\infty} \frac{dr}{\sqrt{g_{yy}(r)g_{xx}(r)}}\right]\partial_{y}J^{t}.
\end{split}
\end{equation}
This relation is same as Fisk's first law, $J^x_{i}=-D_{x_{i}}\partial_{x_{i}}J^{t}$ and the diffusion coefficients which have been calculated from anisotropic horizon in the near horizon limit are,
\begin{equation}
\label{eq:47}
D_{x}=-(\textit{A+iB}){g_{xx}g_{rr}} \sqrt{g_{xx}(r_{0})g_{yy}(r_{0})}\int_{r_{0}}^{\infty} \frac{dr}{\sqrt{g_{xx}(r)g_{yy}(r)}},
\end{equation}
and
\begin{equation}
\label{eq:48}
D_{y}=-(\textit{M+iN}){g_{yy}g_{rr}} \sqrt{g_{yy}(r_{0})g_{yy}(r_{0})}\int_{r_{0}}^{\infty} \frac{dr}{\sqrt{g_{yy}(r)g_{xx}(r)}}.
\end{equation}

As above mentioned, the diffusion coefficient can be complex depending on the form of metric components, and the fact that we took the perturbed vector field to propagate in two dimensions. Although most of static black holes along with the assumption of having one dimensional propagating vector field leads to the real diffusion coefficient. Our result shows the modified diffusion coefficient when we  take the gauge field  propagates in  two dimensions.
As we know, the variation of Einstein-Maxwell action will produce a surface term and the total $x$-directed current on the horizon will be,
\begin{equation}
\label{eq:49}
\begin{split}
J^{x}&=\frac{1}{g_{xx}\sqrt{g_{rr}}}F_{xr}=-(\textit{A+iB}){g_{xx}\sqrt{g_{rr}}} F_{tx}, \\
     &=-(\textit{A+iB}){g_{xx}\sqrt{g_{rr}}} E_{x}\mid_{r_{h}},  \\
     &=\sigma^{xx} E_{x}\mid_{r_{h}}.
\end{split}
 \end{equation}
This relation express $DC$ conductivity on the horizon for anisotropic space times,
\begin{equation}
\label{eq:50}
 \sigma^{xx}=-(\textit{A+iB}){g_{xx}\sqrt{g_{rr}}} .
 \end{equation}
and also we have,
\begin{equation}
\label{eq:51}
\begin{split}
J^{y}&=\frac{1}{g_{yy}\sqrt{g_{rr}}}F_{yr}=-(\textit{M+iN}){g_{yy}\sqrt{g_{rr}}} F_{ty}, \\
     &=-(\textit{M+iN}){g_{yy}\sqrt{g_{rr}}} E_{y}\mid_{r_{h}},  \\
     &=\sigma^{yy} E_{y}\mid_{r_{h}}.
\end{split}
 \end{equation}
 which leads to,
\begin{equation}
\label{eq:52}
 \sigma^{yy}=-(\textit{M+iN}){g_{yy}\sqrt{g_{rr}}} .
 \end{equation}

\section{AdS/CFT duality and deriving transport and  diffusion coefficient }
In this section we try to extract transport coefficient in the hydrodynamic limit of $AdS/CFT$ duality. First of all we calculate retarded Green's function from $AdS/CFT$ duality construction. So an appropriate perturbed vector field must be considered,
\begin{equation}
\label{eq:53}
A_{x}=a_{x}(r) e^{-i\omega t + iq x}.
\end{equation}
The Maxwell equations for this perturbed vector field are,
\begin{equation}
\label{eq:54}
\begin{split}
&\sqrt{\frac{g_{yy}}{g_{xx}}}g^{tt}\partial_{x}\partial_{t}A_{x}=0, \\
&\sqrt{\frac{g_{yy}}{g_{xx}}}g^{rr}\partial_{x}\partial_{r}A_{x}=0, \\
&\sqrt{\frac{g_{yy}}{g_{xx}}}g^{tt}\partial_{t}^2A_{x}+ \partial_{r}\left(\sqrt{\frac{g_{yy}}{g_{xx}}}g^{rr}\partial_{r}A_{x} \right)=0.
\end{split}
\end{equation}
Now by inserting Eq. (53) into the last equation of (54) we can extract dynamical equation of the corresponding perturbed mode. So the corresponding differential equation will be,
\begin{equation}
\label{eq:55}
\partial_{r}\left(\sqrt{\frac{g_{yy}}{g_{xx}}}g^{rr}\partial_{r}a_{x} \right)-\sqrt{\frac{g_{yy}}{g_{xx}}}\frac{\omega^{2}}{g_{tt}}a_{x}=0,
\end{equation}
where  in the $\omega\rightarrow0$ limit, we have following equation,
\begin{equation}
\label{eq:56}
a_{x}(r)=C_{1}\int_{r}^{\infty}g_{rr}(r')\sqrt{\frac{g_{xx}(r')}{g_{yy}(r')}}dr'.
\end{equation}
Now  we suppose the differential equation (55) has the following solution,
\begin{equation}
\label{eq:57}
a_{x}(r)=exp\left(i\alpha\omega\int_{r}^{\infty}g_{rr}(r')\sqrt{\frac{g_{xx}(r')}{g_{yy}(r')}}dr'\right).
\end{equation}
By replacing this solution into the Eq. (55) the parameter $\alpha$ will be,
\begin{equation}
\label{eq:58}
\alpha=\sqrt{-\frac{g_{yy}}{g_{tt}g_{rr}g_{xx}}}.
\end{equation}
Since the $\omega\rightarrow0$ limit is suitable for our purpose, then,
 \begin{equation}
\label{eq:59}
a_{x}(r)=1+i\alpha\omega\int_{r}^{\infty}g_{rr}(r')\sqrt{\frac{g_{xx}(r')}{g_{yy}(r')}}dr'+ O(\omega^{2}).
\end{equation}
From $AdS/CFT$ duality we can insert this solution to the boundary term of the action,
\begin{equation}
\label{eq:60}
\begin{split}
S_{on-shell}&=-\frac{1}{2}\int_{r\rightarrow\infty}   d^{3}x  \sqrt{-g} A_{x}g^{rr}g^{xx} F_{rx}, \\
            &=-\frac{1}{2}\int_{r\rightarrow\infty}   d^{3}x  A_{x} \left(i\alpha\omega \sqrt{-g} \frac{1}{\sqrt{g_{xx}g_{yy}}}\right) A_{x}.
\end{split}
\end{equation}
So the retarded Green's function will be,
\begin{equation}
\label{eq:61}
G^{R}(\omega,q\rightarrow0)= i\alpha\omega\sqrt{-g} \frac{1}{\sqrt{g_{xx}g_{yy}}},
\end{equation}
Now we use Kubo formula and determine $DC$ conductivity as follows,
\begin{equation}
\label{eq:62}
\sigma ^{xx}= \lim_{\omega\rightarrow0} \frac{1}{\omega} ImG^{R}(\omega,0)\mid_{r_{0}}=\alpha  \frac{\sqrt{-g}}{\sqrt{g_{xx}g_{yy}}}
=\frac{\sqrt{-g}}{g_{xx}\sqrt{g_{tt}g_{rr}}}\mid_{r_{0}}.
\end{equation}

The diffusion coefficient is related to the corresponding transport coefficient such that $D=\frac{\sigma^{DC}}{\chi}=\frac{\sigma^{xx}}{\chi}$,
 where  $\chi$ is the charge susceptibility which is completely introduced in Ref.~\cite{ee}.

Now we are going to consider two known examples and study the corresponding parameters. First example which we consider is Einstein-Maxwell-dilaton-axion Model, so we introduce one example of anisotropic black hole which is important in studying holographic linear response theory. We take the following action~\cite{hh},
\begin{equation}
\label{eq:63}
S=\int d^{4} x\sqrt{-g}\left(R+\frac{6}{L^{2}}-2(\nabla\phi)^2-\frac{1}{2}e^{4\phi}\Sigma_{i=1}^{2}(\nabla{\textmd{a}_{i}})^2-e^{-2\phi}F^{2}\right),
\end{equation}
 where $\phi$ is dilaton and $\textmd{a}_{i}$ represents two axion fields and $L$ is the $AdS$ radius. Turning on axion fields causes the momentum relaxation, so the dual field theory will represent anisotropic medium. The appropriate metric which relates to the above action is,
\begin{equation}
\label{eq:64}
ds^2=L^2\left(-r^2 g(r)dt^2+\frac{1}{r^2 g(r)}dr^2+e^{A(r)+B(r)}dx^{2}+e^{A(r)-B(r)}dy^{2}\right).
\end{equation}
Unfortunately the explicit expressions for the metric components are not exist and we can not extract explicit results by our method. Assuming some boundary terms and constraints for this spacetime one can obtain discrete solutions and it will be interesting to calculate transport coefficient as a separate work.

The second example is Einstein-Maxwell-dilaton-axion Model in Massive Gravity.
As we know breaking the spatial translation symmetry leads to the finite $DC$ conductivity. In the holographic model there are several ways which causes to the translational symmetry breaking. In the previous example we introduce linear axion field which imposes momentum relaxation geometry. In this example we add Axionic-Chern-Simons term and because this term leads to the nontrivial Hall conductivity, we take massive gravity theory which will solve the problem. The appropriate action is~\cite{ii},
\begin{equation}
\label{eq:65}
S=\int d^{4} x\sqrt{-g}\left(R+\frac{6}{L^{2}}-2(\nabla\phi)^2-\frac{1}{2}e^{4\phi}(\nabla{\textmd{a}})^2-e^{-2\phi}F^{2}-\textmd{a}\textbf{F}F+p_1[K]+p_2([K]^{2}-[K^2])\right),
\end{equation}
where $p_{1}$ and $p_{2}$ are constant parameters. Also $[K]$ is the trace of square root tensor defined by $K^{\mu}_{\sigma}K^{\sigma_{\nu}}\equiv g^{\mu\sigma}f_{\sigma\nu}$ and $f_{\mu\nu}$ is the reference metric which breaks diffeomorphism in $x,y$ directions,
\begin{equation}
\label{eq:66}
f_{\mu\nu}=diag[0,0,k_{1}^2 H(r)^2,k_{2}^2 H(r)^2].
\end{equation}
The following action can lead to the anisotropic spacetime,
\begin{equation}
\label{eq:67}
ds^2=L^2\left(-r^2 g(r)dt^2+\frac{1}{r^2 g(r)}dr^2+e^{2U_{1}(r)}dx^{2}+e^{2U_{2}(r)}dy^{2}\right).
\end{equation}
Again we emphasize that this model dose not have explicit analytical solution which express metric components in terms of radial coordinate. Also here in order to solve the problem we must impose some constraints and boundary conditions, the corresponding conditions lead us to have the discrete solutions.

\section{Conclusion and outlook }
In this paper, we tried to extract transport coefficients for four dimensional spatially anisotropic static black holes. We have employed two methods to obtain transport coefficients. First of all adapting  from membrane paradigm we have choose stretched horizon as the appropriate surface. Then we perturbed the corresponding spacetime by the vector gauge field which is propagating in two dimensions. Applying Maxwell equations and Bianchi identities we showed that the assumption of having two dimensional propagating gauge field causes modification of the diffusion coefficient. In addition, the obtained results will be complex. So when the metric components identified explicitly the exact solution for the diffusion coefficient can be extracted from (47) and (48) and $DC$ conductivity can be extracted from (50) and (52).

On the other hand, we have reviewed electro-thermal method and used  suitable vector field. In that case, in order to obtain  the $DC$ conductivity one can find retarded Green's function and apply  Kubo formula. So, the diffusion constant can be obtained by $DC$ conductivity and susceptibility.
Then we presented two examples which are best selection for the black holes. Because the explicit form of the metric functions are not given, constraints and boundary conditions yield some discrete solutions. In future work, it may be interesting to find discrete transport coefficients for two pointed spacetimes.

\end{document}